# Tuck's incompressibility function: statistics for zeta zeros and eigenvalues

M V Berry[1] & P Shukla[2]

[1]H H Wills Physics Laboratory, Tyndall Avenue, Bristol BS8 1TL, UK

[2]Department of Physics, Indian Institute of Technology, Kharagpur, India

**Abstract**

For any function $D(x)$ that is real for real $x$, positivity of Tuck's function $Q(x) \equiv D'^2(x)/\left(D'^2(x) - D(x)D''(x)\right)$ is a condition for the absence of complex zeros close to the real axis. Study of the probability distribution $P_N(Q)$, for $D(x)$ with $N$ zeros corresponding to eigenvalues of the Gaussian unitary ensemble (GUE), supports Tuck's observation that large values of $Q$ are very rare for the Riemann zeros. $P_N(Q)$ has singularities at $Q=0$, $Q=1$ and $Q=N$. The moments (averages of $Q^m$) are much smaller for the GUE than for uncorrelated random (Poisson-distributed) zeros. For the Poisson case, the large-$N$ limit of $P_N(Q)$ can be expressed as an integral with infinitely many poles, whose accumulation, requiring regularization with the Lerch transcendent, generates the singularity at $Q=1$, while the large-$Q$ decay is determined by the pole closest to the origin. Determining the large-$N$ limit of $P_N(Q)$ for the GUE seems difficult.

## 1. Introduction

For a function $D(x)$ that is real for real $x$ and analytic in a strip including the real axis, the modulus $|D|$ near the real axis is

$$|D(x+iy)|^2 = D^2(x) + \left(D'^2(x) - D(x)D''(x)\right)y^2 + \cdots \qquad (1.1)$$

$D(x)$ may possess real zeros, but if the coefficient of $y^2$ is positive there will be no complex zeros in the immediate neighbourhood of the real axis. This motivated Tuck [1], in unpublished work, to study this coefficient, normalized by $D'^2(x)$. We find it more convenient to study the reciprocal, namely

$$Q(x) \equiv \frac{D'^2(x)}{D'^2(x) - D(x)D''(x)}. \qquad (1.2)$$

In an investigation of the Riemann zeros, Tuck studied $Q(x)$ numerically for

$$D(x) = x^{1/4} \exp\left(\tfrac{1}{4}\pi x\right) \pi^{-\tfrac{1}{2}ix} \Gamma\left(\tfrac{1}{4} + \tfrac{1}{2}ix\right) \zeta\left(\tfrac{1}{2} + ix\right), \qquad (1.3)$$

which is real on the critical line $x$ real. Computing for $x<30,000,000$, he found no negative values of $Q(x)$, and very few large values: the largest $Q$ he found, for $x$ close to 24,476,747 (figure 1a) was $Q=2.86033$.

Our aims here are twofold. First, to demystify Tuck's observation by showing that it is consistent with modelling the Riemann zeros by eigenvalues in the Gaussian unitary ensemble (GUE) of random matrices [2]. Second, and more generally, to explore $Q(x)$ as an unfamiliar and interesting statistic for the zeros of real functions, for example the characteristic polynomials (secular determinant) of random-matrix theory, where the zeros are the eigenvalues.

To begin, several elementary observations: First, $Q(x)$ is invariant under rescaling of the magnitude of $D$ or of the $x$ axis. Second, $Q(x)$ possesses zeros between pairs of real zeros of $D$, at the critical points where $D'(x)=0$. Third, when $x=x_0$ corresponds to a simple zero of $D(x)$, it follows from (1.2) that $Q(x_0)=1$. Fourth, in a degenerate situation where $N$ zeros coincide at $x=x_0$, so that $D(x) \sim a(x-x_0)^N$, a similar argument gives $Q(x_0)=N$, so large values of $Q$ are associated with near-degeneracies of many zeros of $D(x)$; thus Tuck's function can be regarded as a measure of incompressibility of the zeros. And fifth, positivity of $Q(x)$ does not guarantee the absence of complex zeros that are not close to the real axis; this is illustrated by

$$D(x,a) = \left(x^2 - 1\right)\left(x^2 + a^2\right), \tag{1.4}$$

for which $Q(x)$ is positive for all real $x$ if $a>1$, that is if the separation of the complex zeros $x=\pm ia$ exceeds that of the real zeros $x=\pm 1$. Therefore Tuck's computations for $\zeta$ do not indicate the absence of complex zeros far from the critical line, and should not be interpreted as implying that the Riemann hypothesis is true (for this, $|D(x)|$ must be studied away from the critical line [3]).

We will study $Q(x)$ for functions with $N$ real zeros, the ultimate interest being the limit $N \to \infty$, and we choose the form

$$D_N(x) = \prod_{n=1}^{N}(x - x_n), \tag{1.5}$$

Then an elementary calculation gives (and defining sums $S_{1N}$ and $S_{2N}$)

$$Q_N(x) = \frac{\left(\sum_{n=1}^{N} \frac{1}{(x-x_n)}\right)^2}{\sum_{n=1}^{N} \frac{1}{(x-x_n)^2}} \equiv \frac{S_{1N}^2(x)}{S_{2N}(x)}. \tag{1.6}$$

For the cases we are interested in, this converges if $S_{1N}$ is summed by pairing positive and negative $x_n$. $Q_N(x)$ is obviously positive if all $x_n$ are real, but it can be negative as $x$ passes between a complex-conjugate pair of zeros close to the real axis. If all $x_n$ are real, maxima of $Q(x)$ can never be less than unity (Appendix 1, and see figure 1a).

Our interest will be in the statistics of $Q$, as embodied in the moments

$$M_{mN} = \left\langle (Q_N(x))^m \right\rangle, \tag{1.7}$$

and the probability distribution

$$P_N(Q) = \left\langle \delta(Q - Q_N(x)) \right\rangle. \tag{1.8}$$

The averages $\langle \cdots \rangle$ here are over ranges of $x$.

A trivial nonrandom example, illustrating several of the observations made earlier, is the function

$$D(x) = \sin \pi x, \tag{1.9}$$

whose infinitely many zeros are the integers. From (1.2),

$$Q(x) = \cos^2 \pi x, \tag{1.10}$$

whence (1.8) gives

$$P(Q) = \int_0^1 dx \delta\left(Q - \cos^2 \pi x\right) = \frac{\Theta(1-Q)}{\pi\sqrt{Q(1-Q)}}, \quad (1.11)$$

in which the unit step $\Theta$ denotes the cutoff at $Q=1$.

Figure 1b shows $P(Q)$ for the Riemann zeta function (1.3), over a range including several thousand zeros. This exhibits singularities at $Q=0$ and $Q=1$, that will be described later, and a rapid decay as $Q$ increases, illustrating Tuck's observation that large values of $Q$ are very rare.

In what follows, we will calculate the averages for Poisson-distributed zeros (uncorrelated random numbers) and GUE-distributed eigenvalues; as described in section 2, these ensembles are ergodic so we can replace averaging over $x$ by ensemble averaging. In section 3, we show that $P_N(Q)$ has singularities at $Q=0$ and $Q=1$ (as seen already in figure 1b), and also, for finite $N$, at $Q=N$. We establish the form of the singularities, and illustrate them by calculating $P_N(Q)$ exactly for $N=2$ and $N=3$ for the Poisson and GUE ensembles.

Section 4 contains calculations of the moments $M_{mN}$ for Poisson-distributed zeros, for finite and the large $N$ limit. Section 5 is a calculation of the full distribution $P(Q)$ in the large $N$ limit, for the Poisson case; this turns out to be surprisingly intricate, involving regularization of an infinite accumulation of poles by the Lerch transcendent function. Section 6 is the corresponding calculation of $M_{mN}$ for the GUE; here we are unable to calculate the large-$N$ limit, but display explicit results for $1 \leq N \leq 6$ and $1 \leq m \leq 6$. The GUE moments are systematically smaller than those for the Poisson distribution, reflecting the much more rapid decay of the tail of $P(Q)$ in the GUE case. Given the accuracy with which the GUE reproduces the statistics of the Riemann zeros, this explains why Tuck observed very few large values of $Q$. A full calculation of $P(Q)$ for the GUE case seems difficult; Section 7 contains some remarks about it.

If $D(x)$ is a Gaussian random function (e.g. a Fourier series containing many terms, with random phases), it will typically possess complex zeros; then $P(Q)$ can be negative and is very different from the Poisson and GUE ensembles. The corresponding one-parameter family of $P(Q)$ is calculated in Appendix 2.

We note that in the case where the $x_n$ are the eigenvalues of a matrix $\mathbf{H}$, $Q(x)$ can be written in several alternative forms, for example

$$Q(x) = \frac{\mathrm{Tr}^2(x-\mathbf{H})^{-1}}{\mathrm{Tr}(x-\mathbf{H})^{-2}} = -\left[\frac{\partial}{\partial x}\left[\frac{\partial}{\partial x}\mathrm{Tr}\log(x-\mathbf{H})\right]^{-1}\right]^{-1}. \tag{1.12}$$

## 2. Ensemble averaging

For the ensembles we consider, averaging will eliminate the dependence on $x$ provided $x$ is not close to the smallest or largest zero $x_1$ or $x_N$, and it is convenient to set $x=0$. Thus we study

$$Q_N \equiv Q_N(0) = \frac{\left(\sum_{n=1}^{N}\frac{1}{x_n}\right)^2}{\sum_{n=1}^{N}\frac{1}{x_n^2}} \equiv \frac{S_{1N}^2}{S_{2N}}. \tag{2.1}$$

This quantity is obviously invariant under rescaling of the set of zeros $x_n$.

Figure 2a shows the distribution $P(Q)$ for Poisson-distributed zeros, together with theory to be described later. Figure 2b shows the corresponding graph for GUE-distributed eigenvalues, together with the the distribution of Riemann zeros of figure 1b; evidently the fit is excellent, consistent with known results for other statistics. Note that the decay as $Q$ increases is much faster for the GUE than for Poisson.

As preliminaries, we formulate the two quantities we will be dealing with. First, the moments of the distribution $P_N(Q)$: from (1.7), these are

$$M_{mN} = \left\langle \left( \frac{S_{1N}^2}{S_{2N}} \right)^m \right\rangle. \qquad (2.2)$$

The strategy for getting a calculable form is to express the quantity to be averaged in terms of products of the zeros $x_n$. First, we use the multinomial theorem to simplify the powers of the sum $S_1$:

$$(S_{1N})^{2m} = (2m)! \times \text{coefficient of } t^{2m} \text{ in } \prod_{n=1}^{N} \exp\left(\frac{t}{x_n}\right). \qquad (2.3)$$

Then, to simplify the inverse powers of $S_2$, we write

$$(S_{2N})^{-m} = \frac{2^{1-m}}{\Gamma(m)} \int_0^\infty dz\, z^{2m-1} \exp\left\{-\tfrac{1}{2} z^2 S_{2N}\right\}. \qquad (2.4)$$

Thus the moments are

$$M_{mN} = 2^{1-m} \prod_{n=m}^{2m} n \int_0^\infty dz\, z^{2m-1} B_{mN}(z,t), \qquad (2.5)$$

where

$$B_{mN}(z,t) = \text{coefficient of } t^{2m} \text{ in } \left\langle \prod_{n=1}^{N} \exp\left\{-\frac{z^2}{2x_n^2} + \frac{t}{x_n}\right\} \right\rangle. \quad (2.6)$$

The second quantity is the probability distribution: from (1.8), we have

$$P_N(Q) = \iint dxdy\, \delta\left(Q - \frac{y^2}{x}\right) \langle \delta(y - S_{1N})\delta(x - S_{2N}) \rangle. \quad (2.7)$$

This leads to

$$\begin{aligned}
P_N(Q) &= \frac{1}{(2\pi)^2} \iint dxdy \iint dtdu\, \delta\left(Q - \frac{y^2}{x}\right) \\
&\quad \times \exp\{i(xu+yt)\} \langle \exp\{-i(S_{2N}u + S_{1N}t)\} \rangle \\
&= \frac{1}{4}\left(\frac{i}{\pi}\right)^{3/2} \frac{\partial}{\partial Q}\sqrt{Q} \int_{-\infty}^{\infty} \frac{du}{u^{3/2}} \int_{-\infty}^{\infty} dt \exp\left\{-i\frac{t^2 Q}{4u}\right\} \\
&\quad \times \left\langle \prod_{n=1}^{N} \exp\left\{-i\left(\frac{u}{x_n^2} + \frac{t}{x_n}\right)\right\} \right\rangle \\
&= 2\frac{\partial}{\partial Q}\sqrt{Q}\, \text{Re}\left(\frac{i}{\pi}\right)^{3/2} \int_{0}^{\infty} \frac{d\sigma}{\sigma} \int_{0}^{\infty} d\tau \exp\left\{-i\tfrac{1}{4}\tau^2 Q\right\} \\
&\quad \times \left\langle \prod_{n=1}^{N} \exp\left\{-i\left(\frac{\sigma^2}{x_n^2} + \frac{\tau\sigma}{x_n}\right)\right\} \right\rangle,
\end{aligned} \quad (2.8)$$

after some elementary transformations.

To calculate the averages, we need the joint probability distributions of the $x_n$. For the Poisson case, we take $N$ uncorrelated random numbers whose density is Gauss-distributed:

$$P_{\text{Poisson}}(x_1, \cdots x_N) = \frac{1}{(2\pi)^{N/2}} \exp\left\{-\tfrac{1}{2} \sum_{n=1}^{N} x_n^2\right\}. \tag{2.9}$$

In the GUE case, the distribution is [2]

$$P_{\text{GUE}}(x_1, \cdots x_N) = \frac{1}{A_N} \prod_{\substack{1 \le m, n \le N \\ m < n}} (x_m - x_n)^2 \exp\left\{-\tfrac{1}{2} \sum_{n=1}^{N} x_n^2\right\}, \tag{2.10}$$

where

$$A_N = (2\pi)^{N/2} \prod_{n=1}^{N} n! = (2\pi)^{N/2} \prod_{n=1}^{N} n^{N-n}. \tag{2.11}$$

Alternatively, and as will be discussed in section 8, we could use the circular ensembles [2], in which the zeros lie on the unit circle in the complex plane, but this seems not to lead to any significant simplification in the calculations to follow.

We will make extensive use of the following integral, expressible as modified Bessel functions of half-integer order and also as a finite sum:

$$F(s,z) \equiv \sqrt{\frac{2}{\pi}} \exp(z) \int_0^\infty dx x^{2s} \exp\left\{-\frac{1}{2}\left(x^2 + \frac{z^2}{x^2}\right)\right\}$$

$$= \sqrt{\frac{2}{\pi}} z^{\frac{1}{2}+s} \exp(z) K_{s+\frac{1}{2}}(z) \qquad (2.12)$$

$$= \frac{1}{2^{\sigma(s)} z^{\sigma(s)-s}} \sum_{n=0}^{\sigma(s)} \frac{(2\sigma(s)-n)!}{n!(\sigma(s)-n)!} (2z)^n$$

$$(\sigma(s) = s \ (s \geq 0), \sigma(s) = |s| - 1 \ (s < 0)).$$

### 3. Singularities of $P_N(Q)$, illustrated by $N=2$ and $N=3$

The first of the three singularities is at $Q=0$, where according to (1.2) there are minima, associated with extrema of $D(x)$ (i.e. zeros of $D'(x)$) between zeros. For each such minimum (say at $x=0$),

$$Q(x) \approx \alpha x^2, \qquad (3.1)$$

so

$$P_N(Q) \sim \int dx \delta(Q - \alpha x^2) = \frac{1}{\sqrt{2\alpha Q}} \quad (Q \ll 1). \qquad (3.2)$$

Thus we expect, independently of $N$ and the nature of the statistics (e.g. Poisson or GUE), that $P(Q)$ has an inverse square-root singularity at $Q=0$, and indeed this is evident in all graphs shown here.

At $Q=1$, there is a singularity, also independent of $N$ and the statistics, associated with the fact that $Q=1$ occurs when a zero passes smoothly through $x=0$ for particular members of the ensemble. In the $N$- dimensional space $\{x_1,...x_N\}$, we now show that this corresponds to a saddle if $N \geq 3$. Write (2.1) in the form

$$Q = \frac{(1 + x_N S_{1,N-1})^2}{(1 + x_N^2 S_{2,N-1})}. \tag{3.3}$$

If $N \geq 3$, $S_{1,N-1}$ can vanish because it can contain both positive and negative terms. This can be accommodated by replacing $x_{N-1}$ by the new coordinate $x$, defined by

$$\frac{1}{x_{N-1}} = -S_{1,N-2} + \xi. \tag{3.4}$$

Expansion of (2.1) now gives

$$Q = 1 + 2x_N \xi - x_N^2 \left( (S_{1,N-2})^2 + S_{2,N-2} \right) + \cdots \tag{3.5}$$

The quadratic form in the $(x_N, \xi)$ plane is always indefinite, so there is indeed a saddle at $Q=1$. It is known [4] that this corresponds to a logarithmic singularity of $P(Q)$ (which from (1.8) is an integral along the $Q$ contours in $x$ space). This can be seen directly in the model

$$Q = 1 + x_1^2 - x_2^2, \tag{3.6}$$

where, with a gaussian convergence factor for the contours far from the saddle,

$$P(Q) = \frac{\varepsilon}{\pi} \int_{-\infty}^{\infty} dx_1 \int_{-\infty}^{\infty} dx_2 \, \delta\left(Q - 1 - x_1^2 + x_2^2\right) \exp\left\{-\varepsilon^2\left(x_1^2 + x_2^2\right)\right\}$$

$$= \frac{1}{\pi} K_0\left(\varepsilon |Q-1|\right) = \frac{-\log|Q-1|}{\pi} + \cdots \quad (|Q-1| << 1) \tag{3.7}$$

where … denotes terms in $\log\varepsilon$ and powers of $|Q-1|$. Thus for $N \geq 3$ we expect a logarithmic singularity at $Q=1$, and this is evident in all graphs shown here.

At $Q=N$, there is a maximum where all $x_n$ coincide. To show this, we use coordinates $x_1$ and $\xi_2...\xi_N$ defined by

$$x_n = (1 + \xi_n) x_1, \quad n = 2, 3 \ldots N, \tag{3.8}$$

and expand about $\xi_n = 0$, noting that $Q$ is independent of $x_1$:

$$Q = \frac{\left(1 + \sum_{n=2}^{N}(1+\xi_n)^{-1}\right)^2}{\left(1 + \sum_{n=2}^{N}(1+\xi_n)^{-2}\right)} = N - \left(\sum_{n=2}^{N}\xi_n^2 - \frac{1}{N}\left(\sum_{n=2}^{N}\xi_n\right)^2\right) + \cdots \tag{3.9}$$

The quadratic form is always negative, confirming that $Q$ has an absolute maximum where all zeros coincide. To determine the form of $P(Q)$ near $Q=N$, we use $N-1$ dimensional polar coordinates in $\xi$ space. Only the radial coordinate $\rho$ is important, and now the result depends not only on $N$ but also on the statistics.

For Poisson statistics the ensemble average (2.9) gives the integral

$$P_N(Q) \sim \int_0^\infty d\rho \rho^{N-2} \delta(Q - N + \rho^2) \sim (N-Q)^{(N-3)/2} \Theta(N-Q), \quad (3.10)$$
$$(N - Q \ll 1)$$

describing the the cutoff at $Q=N$. For GUE statistics, (2.10) gives the extra factor

$$\prod_{\substack{1 \le m,n \le N \\ m<n}} (x_m - x_n)^2 = x_1^{N(N-1)} \prod_{k=2}^{N} (\xi_k)^2 \prod_{\substack{2 \le m,n \le N \\ m<n}} (\xi_m - \xi_n)^2 \quad (3.11)$$
$$\propto \rho^{N(N-1)},$$

and an analogous argument analogous to that leading to (3.10) gives

$$P_N(Q) \sim (N-Q)^{(N^2-3)/2} \Theta(N-Q) \quad (N - Q \ll 1). \quad (3.12)$$

To illustrate these singularities, we consider first the case $N=2$, for which (2.1) gives

$$Q = \frac{(x_1 + x_2)^2}{(x_1^2 + x_2^2)}. \quad (3.13)$$

In polar coordinates, the radial variable does not appear, and $P(Q)$ is an integral over just the angular coordinate $\phi$.

For Poisson statistics,

$$P_{2\text{Poisson}}(Q) = \frac{1}{2\pi} \int_0^{2\pi} d\phi\, \delta\!\left(Q - (\cos\phi + \sin\phi)^2\right)$$

$$= \frac{1}{2\pi} \int_0^{2\pi} d\phi\, \delta\!\left(Q - 2\cos\!\left(\phi - \tfrac{1}{4}\pi\right)^2\right) \quad (3.14)$$

$$= \frac{\Theta(2-Q)}{\pi\sqrt{Q(2-Q)}}.$$

At $Q=0$, this has the expected $1/\sqrt{Q}$ singularity (3.2), and at $Q=2$ it has the singularity is $1/\sqrt{(2-Q)}$, as predicted by (3.10) for $N=2$; figure 3a illustrates this behaviour.

For GUE statistics, there is the additional repulsion factor as in (2.10) and (3.11), so the integral is

$$P_{2\text{GUE}}(Q) = \frac{1}{2\pi} \int_0^{2\pi} d\phi\, (\cos\phi - \sin\phi)^2 \delta\!\left(Q - (\cos\phi + \sin\phi)^2\right)$$

$$= \frac{1}{\pi} \int_0^{2\pi} d\phi\, \cos\!\left(\phi + \tfrac{1}{4}\pi\right)^2 \delta\!\left(Q - 2\cos\!\left(\phi - \tfrac{1}{4}\pi\right)^2\right) \quad (3.15)$$

$$= \frac{\sqrt{2-Q}}{\pi\sqrt{Q}} \Theta(2-Q).$$

At $Q=0$, this has the expected $1/\sqrt{Q}$ singularity (3.2), and at $Q=2$ the singularity is the weaker $\sqrt{(2-Q)}$ as predicted by (3.12) for $N=2$; figure 3b illustrates this behaviour.

For $N=3$, we choose coordinates

$$\{x_1, x_2, x_3\} = \left\{x_1, x_1 r\sqrt{2}\cos\!\left(\phi + \tfrac{1}{4}\pi\right), x_1 r\sqrt{2}\sin\!\left(\phi + \tfrac{1}{4}\pi\right)\right\}, \quad (3.16)$$

$$Q = Q(r,\phi) = \frac{(x_1 x_2 + x_1 x_3 + x_2 x_3)^2}{\left(x_1^2 x_2^2 + x_1^2 x_3^2 + x_2^2 x_3^2\right)} = \frac{(2\cos\phi + r\cos 2\phi)^2}{\left(2 + r^2 \cos^2 2\phi\right)}. \quad (3.17)$$

After evaluating the integral over $x_1$ and using symmetry in $\phi$ the distribution $P_3(Q)$ in the Poisson case becomes

$$P_{3,\text{Poisson}}(Q) = \frac{2}{\pi} \int_0^\pi d\phi \int_0^\infty dr\, r \frac{\delta(Q - Q(r,\phi))}{\left(1 + 2r^2\right)^{3/2}}. \quad (3.18)$$

The $\delta$ function selects the values

$$r_{\pm}(\phi,Q) = \frac{2\left(\cos\phi \pm \sqrt{Q\left(\cos^2\phi - \tfrac{1}{2}(Q-1)\right)}\right)}{(Q-1)\cos 2\phi}, \quad (3.19)$$

which contribute to the integral over $\phi$ whenever they are real and positive, giving

$$P_{3,\text{Poisson}}(Q) = \frac{2}{\pi} \int_0^\pi d\phi \sum_{\substack{\text{real positive}\\ \text{roots } r_{\pm}(\phi,Q)}} \frac{r}{\left(1 + 2r^2\right)^{3/2} |\partial_r Q(r,\phi)|}. \quad (3.20)$$

This is shown in figure 4a, superposed on the distribution numerically computed from sample sets of 3 zeros. The singularity at $Q=3$ is a step discontinuity (of height $P_3(3)=1/6$), as predicted by (3.10) for $N=3$.

For the GUE, the calculation is the same, except for the additional factor

$$(x_1 - x_2)^2 (x_1 - x_3)^2 (x_2 - x_3)^2$$
$$= 4x_1^6 r^2 \sin^2 \phi \left(1 - 2r\cos\phi + r^2 \cos 2\phi\right)^2 . \tag{3.21}$$

Incorporating this into the integration over $x_1$, along with the GUE normalization factor (2.11), leads to

$$P_{3,\text{GUE}}(Q) = \frac{70}{\pi} \int_0^\pi d\phi \sin^2 \phi \sum_{\substack{\text{real positive} \\ \text{roots } r_\pm(\phi)}} \frac{r^3 \left(1 - 2r\cos\phi + r^2 \cos 2\phi\right)^2}{\left(1 + 2r^2\right)^{9/2} \partial_r Q(r,\phi)}. \tag{3.22}$$

This is shown in figure 4b, superposed on the distribution numerically computed from sample sets of 3x3 matrices. As predicted by (3.12) for $N=3$, the singularity at $Q=3$ is weaker than for the Poisson case: $P_{3,\text{GUE}}(Q)$ vanishes as $c(N-Q)^3$ (the coefficient is small: $c=35/2592=0.0135...$).

### 4. Moments of $Q$: Poisson-distributed levels

The Poisson moments are given by (2.5) and (2.6), in which the average, according to the distribution (2.9) and after expanding in powers of $t$ and using the integral (2.12), is

$$\left\langle \prod_{n=1}^N \exp\left\{-\frac{z^2}{2x_n^2} + \frac{t}{x_n}\right\} \right\rangle = \left(\frac{1}{\sqrt{2\pi}} \int_{-\infty}^\infty dx \exp\left\{-\frac{z^2}{2x^2} + \frac{t}{x}\right\}\right)^N$$
$$= \exp(-Nz) \left(\sum_{r=0}^\infty \frac{t^{2r}}{(2r)!} F(-r,z)\right)^N . \tag{4.1}$$

Thus in (2.6)

$$B_{mN}(z) = \exp(-Nz) \times \text{coefficient of } t^{2m} \text{ in } \left( \sum_{r=0}^{\infty} \frac{t^{2r}}{(2r)!} F(-r,z) \right)^N. \quad (4.2)$$

With the series in (2.12) for $F(s,z)$, the integrals over $z$ are elementary and the computations are easily automated using Mathematica$^{TM}$ since the upper limit of the sum over $r$ can be taken as $2m$. Results for the first few $m$ are shown in Table 1.

| $m$ | $M_{mN}$ |
|---|---|
| 1 | 1 |
| 2 | $2 - N^{-1}$ |
| 3 | $6 - 9N^{-1} + 4N^{-2}$ |
| 4 | $\frac{74}{3} - \frac{149}{2}N^{-1} + \frac{509}{6}N^{-2} - 34N^{-3}$ |
| 5 | $130 - \frac{1315}{2}N^{-1} + \frac{2745}{2}N^{-2} - 1340N^{-3} + 496N^{-4}$ |
| 6 | $\frac{4186}{5} - \frac{254811}{40}N^{-1} + \frac{838833}{40}N^{-2} - \frac{723531}{20}N^{-3} + \frac{158979}{5}N^{-4} - 11056N^{-5}$ |
| 7 | $\rightarrow \frac{95\,578}{15}$ as $N \rightarrow \infty$ |
| 8 | $\rightarrow \frac{391698}{7}$ as $N \rightarrow \infty$ |
| 9 | $\rightarrow \frac{19\,492\,422}{35}$ as $N \rightarrow \infty$ |
| 10 | $\rightarrow \frac{390\,290\,014}{63}$ as $N \rightarrow \infty$ |
| 11 | $\rightarrow \frac{380\,830\,494}{5}$ as $N \rightarrow \infty$ |

Table 1. Moments $\langle Q^m \rangle$ for Poisson-distributed zeros

We are interested in the limit $N \rightarrow \infty$. To get a more direct expression for this, we employ the scalings

$$z \rightarrow z/N, \quad t \rightarrow tz/N, \quad (4.3)$$

which transforms (4.1) into

$$\left\langle \prod_{n=1}^{N} \exp\left\{-\frac{z^2}{2N^2 x_n^2} + \frac{tz}{Nx_n}\right\} \right\rangle$$
$$= \exp\left(-z + N\log\left[1 + \sum_{r=1}^{\infty} \frac{1}{(2r)!}\left(\frac{tz}{N}\right)^{2r} F\left(-r, \frac{z}{N}\right)\right]\right). \tag{4.4}$$

For large $N$, only the leading term in the series (2.12) for $F$ survives, so

$$\left\langle \prod_{n=1}^{N} \exp\left\{-\frac{z^2}{2N^2 x_n^2} + \frac{tz}{Nx_n}\right\} \right\rangle$$
$$\to \exp\left\{-z + N\log\left[1 + \frac{z}{N}\sum_{r=1}^{\infty} \frac{(t^2/2)^r}{r!(2r-1)}\right]\right\} \tag{4.5}$$
$$\to \exp\left\{-z\left[\exp\left(\tfrac{1}{2}t^2\right) - \sqrt{\tfrac{\pi}{2}}\, t\, \mathrm{erfi}\left(\frac{t}{\sqrt{2}}\right)\right]\right\},$$

in which erfi is the error function of imaginary argument [5]. Thus (2.5) and (2.6) become

$$M_{mN\to\infty} \equiv M_m = 2^{1-m} \prod_{n=m}^{2m} n \times \text{coefficient of } t^{2m} \text{ in}$$
$$\int_0^\infty \frac{dz}{z} \exp\left\{-z\left[\exp\left(\tfrac{1}{2}t^2\right) - \sqrt{\tfrac{\pi}{2}}\, t\, \mathrm{erfi}\left(\frac{t}{\sqrt{2}}\right)\right]\right\}. \tag{4.6}$$

The integral as written does not converge, but can be regularized with a lower limit $\varepsilon$ and the approximation of the exponential integral [6] for small argument, namely

$$\int_\varepsilon^\infty \frac{dz}{z} \exp(-zg(t)) = E_1(\varepsilon g(t)) \to -\log\varepsilon - \gamma - \log g(t). \tag{4.7}$$

The infinite term $-\log\varepsilon$ does not contribute because it is independent of $t$, as is $\gamma$, leading to

$$M_m = \prod_{n=m}^{2m} n 2^{1-m} \times \text{coefficient of } t^{2m} \text{ in}$$
$$-\log\left\{\exp\left(\tfrac{1}{2}t^2\right) - \sqrt{\tfrac{\pi}{2}}\, t\,\text{erfi}\left(\tfrac{t}{\sqrt{2}}\right)\right\}. \tag{4.8}$$

From this it is easy to calculate many moments. An exact explicit formula has proved elusive, but we can find the large-$m$ asymptotics using Darboux's principle of the nearest singularity [7, 8]. There are two nearest singularities, namely zeros of the argument of the logarithm, at $\pm t_c$, where

$$t_c = 1.3069297... \tag{4.9}$$

The replacement

$$\log\left\{\exp\left(\tfrac{1}{2}t^2\right) - \sqrt{\tfrac{\pi}{2}}\, t\,\text{erfi}\left(\tfrac{t}{\sqrt{2}}\right)\right\} \to \log(t - t_c) + \text{constant} \tag{4.10}$$

leads to

$$M_m \approx \frac{2}{\sqrt{\pi}}\left(m - \tfrac{1}{2}\right)!\left(\frac{2}{t_c^2}\right)^m \quad \text{for } m \gg 1. \tag{4.11}$$

This formula is astonishingly accurate: it approximates the factorially growing moments to better than 1% for $m>2$, and the error decreases exponentially with $m$.

## 5. $P_N(Q)$ in large $N$ limit: Poisson-distributed zeros

We start from (2.8), with the scaling $\sigma \to \sigma/N$, and use the large-$N$ average (4.5) with the replacements

$$z = \sigma\sqrt{2\mathrm{i}}, \quad t = -\frac{\tau}{\sqrt{2}}\exp\left(\tfrac{1}{4}\mathrm{i}\pi\right), \tag{5.1}$$

so that

$$\left\langle \prod_{n=1}^{N} \exp\left\{-\mathrm{i}\left(\frac{\sigma^2}{N^2 x_n^2} + \frac{\tau\sigma}{N x_n}\right)\right\}\right\rangle \to \exp\{-\sigma\chi(\tau)\}, \tag{5.2}$$

where

$$\chi(\tau) = \sqrt{2}\exp\left\{\tfrac{1}{4}\mathrm{i}(\tau^2 + \pi)\right\} + \tau\sqrt{\frac{\pi}{2}}\mathrm{erf}\left(\frac{\tau}{2}\exp\left(-\tfrac{1}{4}\mathrm{i}\pi\right)\right). \tag{5.3}$$

Thus the large-$N$ distribution becomes

$$P_{N\to\infty}(Q) \equiv P(Q)$$
$$= \frac{2}{\pi^{3/2}}\frac{\partial}{\partial Q}\sqrt{Q}\,\mathrm{Re}\int_0^\infty \mathrm{d}\tau \int_0^\infty \frac{\mathrm{d}\sigma}{\sigma}\exp\left\{\tfrac{\mathrm{i}}{4}\left(3\pi - \mathrm{i}\tau^2 Q\right) - \sigma\chi(\tau)\right\}. \tag{5.4}$$

For the integral over $\sigma$ we again use the regularization (4.7). The constant terms lead to a purely imaginary contribution after integrating over $\tau$, so

$$P(Q) = \frac{2}{\pi^{3/2}}\frac{\partial}{\partial Q}\sqrt{Q}\,\mathrm{Re}\int_0^\infty \mathrm{d}\tau\,\exp\left\{\tfrac{\mathrm{i}}{4}\left(3\pi - \mathrm{i}\tau^2 Q\right)\right\}\log\chi(\tau). \tag{5.5}$$

Integrating by parts gives the main result of this section:

$$P(Q) = -\frac{2}{\pi}\frac{\partial}{\partial Q}\text{Im}\int_0^\infty d\tau \,\text{erfc}\left\{\exp\left(\tfrac{1}{4}i\pi\right)\frac{\tau\sqrt{Q}}{2}\right\}\frac{\partial}{\partial \tau}\log\chi(\tau)$$

$$= -\frac{1}{\pi^{3/2}\sqrt{Q}}\text{Im}\exp\left(\tfrac{1}{4}i\pi\right)\int_0^\infty d\tau\,\tau\exp\left(-\tfrac{1}{4}i\tau^2 Q\right)\frac{\partial}{\partial \tau}\log\chi(\tau).$$

(5.6)

To compute this integral, and to understand its large-$Q$ decay and how it contains the logarithmic singularity at $Q=1$, we need the asymptotics of $\log\chi(\tau)$:

$$\frac{\partial}{\partial\tau}\log\chi(\tau) \approx \frac{1}{\tau} - \frac{2}{\tau^2\sqrt{\pi}}\exp\left(\tfrac{1}{4}i(\tau^2+\pi)\right) \text{ for } \tau\gg 1. \qquad (5.7)$$

This shows that when $Q<1$, (5.6) converges if the $\tau$ contour is deformed into the upper half-plane. Then we can replace

$$\frac{\partial}{\partial\tau}\log\chi(\tau) \to \frac{\partial}{\partial\tau}\log\chi(\tau) - \frac{1}{\tau}, \qquad (5.8)$$

because $\exp(i\pi/4)$ times the integral over $\tau$ integral is real, and numerical computation is unproblematic.

If $Q>1$, the situation is more tricky. The integral in (5.6) converges if the $\tau$ contour is deformed into the lower half-plane, but then it inevitably encounters infinitely many poles $\tau_j$ of $\partial\log\chi/\partial\tau$ (zeros of $\chi(\tau)$), located close to the real axis (figure 5). The contributions of these poles must be included. Their approximate location is, from (5.7),

$$\tau_0 = (1-i)\tau_c, \quad \tau_{j>0} = u_j - iv_j, \text{ where}$$
$$u_j \approx 2\sqrt{(j+\tfrac{1}{8})\pi}, \quad v_j \approx \frac{2}{u_j}\log\left(\tfrac{1}{4}\sqrt{\pi}u_j^3\right), \qquad (5.9)$$

where $\tau_c$ is given by (4.9). Thus for $Q>1$,

$$P(Q) = -\frac{1}{\pi^{3/2}\sqrt{Q}} \operatorname{Im} \exp\left(\tfrac{1}{4}i\pi\right) \times$$

$$\int_0^{\infty \exp(-i\pi\alpha)} d\tau \; \tau \exp\left(-\tfrac{1}{4}i\tau^2 Q\right) \frac{\partial}{\partial \tau} \log \chi(\tau) - 2\pi i \sum_{j(\alpha)}^{\infty} \tau_j \exp\left(-\tfrac{1}{4}i\tau_j^2 Q\right), \quad (5.10)$$

where $j(\alpha)$ denotes the pole closest to the origin that the contour has crossed in its deformation away from the real axis.

It follows from (5.9) that the sum over poles is singular as $Q \to 1$. However, the tail of the sum can be regularized using the Lerch transcendent [5, 9], defined by

$$\Phi(z,s,a) = \sum_{k=0}^{\infty} \frac{z^k}{(k+a)^s}. \quad (5.11)$$

The regularization is

$$\sum_J^{\infty} \tau_j \exp\left(-\tfrac{1}{4}i\tau_j^2 Q\right)$$

$$\approx \frac{2\sqrt{2\pi}}{(4\pi^2\sqrt{2})^Q} \exp\left(-\tfrac{1}{4}i\pi Q\right) \sum_{J \gg 1}^{\infty} \frac{\exp(-2\pi i Q j)}{\left(j + \tfrac{1}{8}\right)^{(3Q-1)/2}}$$

$$= \frac{2\sqrt{2\pi}}{(4\pi^2\sqrt{2})^Q} \exp\left(-\tfrac{1}{4}i\pi Q\right) \quad (5.12)$$

$$\times \left( \exp(-2\pi i Q) \Phi\left(\exp(-2\pi i Q), \tfrac{1}{2}(3Q-1), \tfrac{9}{8}\right) - \sum_1^{J-1} \frac{\exp(-2\pi i Q j)}{\left(j + \tfrac{1}{8}\right)^{(3Q-1)/2}} \right).$$

With this, numerical computation is straightforward, and leads to the theoretical curve in figure 2a, agreeing perfectly with the simulations.

Asymptotic analysis, not given here, shows that the Lerch transcendent, representing the collective contribution of the poles $\tau_{j\gg1}$ far from the origin, contains the logarithmic singularity at $Q=1$. The decay of $P(Q)$ for $Q\gg1$ is governed by the pole $\tau_0$ closest to the origin, whose contribution is

$$P(Q) \approx -\frac{1}{\sqrt{\pi Q}} \operatorname{Im} \tau_0 \exp\left(-\tfrac{1}{4}\mathrm{i}\left(\tau_0^2 Q + \pi\right)\right)$$

$$= \sqrt{\frac{2}{\pi Q}}\, \tau_c \exp\left(-\tfrac{1}{2}\tau_c^2 Q\right) \propto \frac{\exp(-0.8540 Q)}{\sqrt{Q}}, \quad (5.13)$$

in conformity with the formula (4.11) for the large-$m$ moments, and with the simulation (figure 2a).

## 6. Moments of $Q$: GUE-distributed levels

To evaluate the average in the moment formulas (2.5-2.6), including the GUE factor $(x_m-x_n)^2$ in (2.10), it is convenient to separate the different $x_n$ using the relation [2]

$$\prod_{k=1}^{N} f(x_k) \prod_{\substack{1\le m,n\le N \\ m<n}}(x_m - x_n)^2 = \det\left(f(x_k)\sum_{j=1}^{N} x_j^{k+l-2}\right), \quad (6.1)$$

where the indices $k$ and $l$ in the determinant run from 1 to $N$, which implies [2]

$$\prod_{k=1}^{N} \int_{-\infty}^{\infty} dx_k f(x_k) \prod_{\substack{1 \le m,n \le N \\ m<n}} (x_m - x_n)^2 = N! \det\left( \int_{-\infty}^{\infty} dx f(x) x^{k+l-2} \right). \quad (6.2)$$

For the average in (2.6), this gives, on using (2.10)

$$\left\langle \prod_{n=1}^{N} \exp\left\{ -\frac{z^2}{2x_n^2} + \frac{t}{x_n} \right\} \right\rangle = \frac{N!}{A_N} \det_N G_{kl}, \quad (6.3)$$

in which the matrix elements are

$$G_{kl} = \frac{1}{\sqrt{2\pi}} \int_{-\infty}^{\infty} dx\, x^{k+l-2} \exp\left\{ -\frac{1}{2}\left( x^2 + \frac{z^2}{x^2} \right) + \frac{t}{x} \right\} \quad (1 \le k,l \le N). \quad (6.4)$$

Expanding in powers of $t$, and using (2.12), gives the explicit formulas

$$\begin{aligned} G_{kl} &= \exp(-z) \sum_{r=0}^{\infty} \frac{t^{2r}}{(2r)!} F\left(z, \tfrac{1}{2}(k+l-2-2r)\right) \quad (k+l \text{ even}) \\ &= \exp(-z) t \sum_{r=0}^{\infty} \frac{t^{2r}}{(2r+1)!} F\left(z, \tfrac{1}{2}(k+l-3-2r)\right) \quad (k+l \text{ odd}). \end{aligned} \quad (6.5)$$

Then the moments $M_{mN}$ are given by (2.5), with, from (2.6),

$$B_{mN}(z,t) = \frac{N!}{A_N} \times \text{coefficient of } t^{2m} \text{ in } \det_N G_{kl}. \quad (6.6)$$

As in the Poisson case, the calculation can be automated using Mathematica, since the upper limit of the sum over $r$ can be taken as $2m$. We have not been able to calculate any of the moments $M_{mN}$ for general $N$, but we can compute them for any chosen values of $m$ and $N$. The lowest few moments are shown in Table 2.

| $M_{mN}$ | $N = 2$ | $N = 3$ | $N = 4$ | $N = 5$ | $N = 6$ |
|---|---|---|---|---|---|
| $m = 1$ | $\frac{1}{2}$ | $\frac{19}{27} = 0.704$ | $\frac{137}{256} = 0.535$ | $\frac{40\,499}{62\,500} = 0.648$ | $\frac{2\,756\,617}{5\,038\,848} = 0.547$ |
| $m = 2$ | $\frac{1}{2}$ | $\frac{17}{27} = 0.630$ | $\frac{1}{2}$ | $\frac{179\,189}{312\,500} = 0.573$ | $\frac{1}{2}$ |
| $m = 3$ | $\frac{5}{8}$ | $\frac{50}{81} = 0.617$ | $\frac{569}{1024} = 0.556$ | $\frac{1\,771\,531}{3\,125\,000} = 0.567$ | $\frac{5\,401\,439}{10\,077\,696} = 0.536$ |
| $m = 4$ | $\frac{7}{8}$ | $\frac{476}{729} = 0.653$ | $\frac{349}{512} = 0.682$ | $\frac{9\,515\,959}{15\,625\,000} = 0.609$ | $\frac{57\,277\,843}{90\,699\,264} = 0.632$ |
| $m = 5$ | $\frac{21}{16}$ | $\frac{2201}{2916} = 0.755$ | $\frac{7357}{8192} = 0.898$ | $\frac{44\,506\,429}{62\,500\,000} = 0.712$ | $\frac{72\,554\,701}{90\,699\,264} = 0.800$ |
| $m = 6$ | $\frac{33}{16}$ | $\frac{8561}{8748} = 0.979$ | $\frac{5135}{4096} = 1.254$ | $\frac{1\,431\,560\,627}{1\,562\,500\,000} = 0.916$ | $\frac{586\,857\,667}{544\,195\,584} = 1.078$ |

Table 2. Moments $\langle Q^m \rangle$ for GUE-distributed eigenvalues

The main feature is the smallness of the moments, especially striking when compared (Table 3) with those for Poisson-distributed zeros.

| $m$ | 1 | 2 | 3 | 4 | 5 | 6 |
|---|---|---|---|---|---|---|
| $\dfrac{M_{m6}^{\text{GUE}}}{M_{m6}^{\text{Poisson}}}$ | 0.5471 | 0.2727 | 0.1162 | 0.0437 | 0.0152 | 0.0050 |

Table 3. Ratio of GUE moments to Poisson moments.

All the moments are of order unity, and of course they satisfy momemt inequalities, e..g. [10] $M_2^2 \leq M_1 M_3$. More important, they increase very slowly with $m$, reflecting the rarity of large values of $Q$, as observed by Tuck for the Riemann zeros.

## 7. $P_N(Q)$ in large $N$ limit: GUE-distributed eigenvalues

We have not succeeded in finding an explicit formula for the large-$N$ limit $P(Q)$ of $P_N(Q)$ for GUE-distributed eigenvalues, analogous to (5.6) in the Poisson case. The difficulty is in determining the large-$N$ limit of the determinant in the average (2.8), essentially the same as $det_N G_{kl}$ for the moments (equations 6.3 and 6.4). Using (2.12), we can express the matrix elements (6.4-6.5) as power series in $z$, but in contrast to the Poisson case the large-$N$ limit is not determined by the lowest-order term, and we are not sure what $N$ scaling to apply (analogous to (4.3) or (5.1)) to reach the limit. The problem seems hard; an analogy is the level spacings distribution, which was easy to find for $N=2$ but is difficult in the large-$N$ limit.

We cannot even be certain that the limit exists. However, our simulations strongly suggest that it does: as $N$ increases, the distributions quickly renormalize to a constant form: even $P_3(Q)$ (figure 4b) looks qualitatively similar to $P_{20}(Q)$ (figure 2b), and $P_{10}(Q)$ (not shown here) is visually indistinguishable from $P_{20}(Q)$. Assuming the limit does exist, the moments in Table 2, and the simulations, indicate that the decay of the probability tail for large $Q$ is very rapid. If the decay is exponential, the coefficient of $Q$ in the exponent must be much larger than the value $\tau_c^2/2=0.8540$ for the Poisson case (equation (5.13)); numerical simulations suggest when fitting by $\exp(-AQ)/\sqrt{Q}$ the coefficient $A$ is approximately 5.6, but this value should not be taken too seriously because it is based on numerics between $Q=1$ and $Q\sim 2$, which is hardly large $Q$.

Although in principle $Q$ could reach a value $N$ for $N$ eigenvalues, corresponding to a degeneracy for which they all coincide, this is so improbable that in our many simulations we always found $Q<3$. This is consistent with Tuck's observation for the Riemann zeros, where the largest value was $Q=2.86$.

## 8. Concluding remarks

The function $Q(x)$ (equation (1.3)) devised by Tuck, and its statistical counterpart $Q_N$ (equation (2.1)) are sensitive indicators of the compressibility of the set of zeros. In particular, the decay of $P_N(Q)$ for large $Q$, or equivalently the growth of the moments $M_{mN}$ as $m$ increases, reflects the decreasing probability of near-degeneracies of many zeros. In this respect, the tail of $P_N(Q)$ is analogous to the negative moments of the secular polynomial (spectral determinant) near the real axis [11-13]; associated with these are exponents related to eigenvalue coalescences. The advantage of $Q$ is that it is scale-invariant.

Concerning the Riemann zeros, Tuck's observation of the rarity of large values of $Q$ turns out to be an unexpected consequence of the known fact [2] that the short-range statistics of the zeros are accurately described by those of the GUE.

We have explored Poisson-distributed zeros and GUE-distributed eigenvalues. It is natural to ask about the analogous statistics for the Gaussian orthogonal ensemble (GOE) and the Gaussian symplectic ensemble (GSE). We have carried out simulatons for these cases. Unsurprisingly, the tail for GOE lies between those of the Poisson and GUE cases, and the tail for the GSE decays fastest of all (we found no $Q$ values larger than 2).

As mentioned in section 2, the entire analysis could have been carries out on the unit circle rather than the real line. Instead of (1.5), we could have studied the periodic function

$$D_N(x) = \prod_{n=1}^{N} \sin(2\pi(x - x_n)), \qquad (8.1)$$

with antipodal pairs of zeros at $x_n$ and $x_n+1/2$. And instead of (1.6) we would have

$$Q_N(x) = \frac{\left(\sum_{n=1}^{N} \cot(\pi(x - x_N))\right)^2}{\sum_{n=1}^{N} \csc^2(\pi(x - x_N))}. \qquad (8.2)$$

Instead of the Poisson statistics, we would have random zeros on the circle, and instead of the GUE we would have the circular unitary ensemble (CUE) [2] in which eigenvalue differences $(x_m - x_n)^2$ are replaced by chord lengths $|\exp(i\, x_m) - \exp(i\, x_n)|^2$. Some details of the analysis are different: incomplete gamma functions replace the $K$ Bessel functions in (2.12). But the unsolved problem of determining the large-$N$ limit $P(Q)$ seems just as difficult with the CUE as with the GUE.

**Acknowledgments**

MVB thanks Professor E O Tuck for discussions and generously supplying unpublished calculations, and Professor Jon Keating for helpful suggestions. Both authors thank ICTP and SISSA, Trieste, for support while this work was completed.

**Appendix 1. Maxima of $Q_N(x)$ cannot be less than unity.**

From (1.6), and in an obvious notation,

$$\frac{\partial S_{kN}(x)}{\partial x} = -S_{k+1N}(x). \tag{A.1}$$

Thus

$$\frac{\partial Q(x)}{\partial x} = \frac{2S_{1N}^2 S_{3N}}{S_{2N}^2} - 2S_{1N}, \tag{A.2}$$

which has minima when $S_{1N}=0$, where $Q_N(x)=0$, and maxima when

$$S_{1N} S_{3N} = S_{2N}^2 . \tag{A.3}$$

It follows that the maxima of $Q_N(x)$ take the values

$$Q_{N\max} = \frac{S_{2N}^3}{S_{3N}^2}. \tag{A.4}$$

The fact that this can never be less than unity is an immediate consequence of Hölder's inequality,

$$\left(\sum_1^N a_n^p\right)^{1/p} \left(\sum_1^N b_n^q\right)^{1/q} \geq \sum_1^N a_n b_n, \tag{A.5}$$

with

$$a_n = \frac{1}{(x-x_n)^3}, p = \frac{2}{3}, b_n = 1, q = \infty \tag{A.6}$$

(in this application, the usual restriction that the $a_n$ be non-negative is unnecessary).

## Appendix 2. Zeros of Gaussian random functions

The ensemble of functions can be defined as

$$D(x) = \sum_{\kappa} a_{\kappa} \cos(\kappa x + \phi_{\kappa}), \tag{A.7}$$

with many Fourier components $\kappa$, fixed real amplitudes $a_{\kappa}$, and random phases $\phi_{\kappa}$. Define the quantity

$$Z(x) \equiv D(x)D''(x), \tag{A.8}$$

which for Gaussian random functions is statistically independent of $D'(x)$. Then from the definition (1.2) the distribution of $Q$ is

$$P(Q) = \int_{-\infty}^{\infty} dD' \int_{-\infty}^{\infty} dZ P_{D'}(D') P_Z(Z) \delta\left(Q - 1\bigg/\left(1 - \frac{Z}{D'^2}\right)\right), \tag{A.9}$$

In this integral, the distributions $P$ in the integral refer to the indicated quantities. Relevant averages are

$$\left\langle D^2 \right\rangle = \tfrac{1}{2}\sum a_{\kappa}^2 \equiv \kappa_0, \quad \left\langle D'^2 \right\rangle = \tfrac{1}{2}\sum \kappa^2 a_{\kappa}^2 \equiv \kappa_2$$
$$\left\langle D''^2 \right\rangle = \tfrac{1}{2}\sum \kappa^4 a_{\kappa}^2 \equiv \kappa_4, \quad \left\langle DD'' \right\rangle = -\tfrac{1}{2}\sum \kappa^2 a_{\kappa}^2 = -\kappa_2, \tag{A.10}$$

and enter the basic distributions

$$P_{D'}(D') = \frac{1}{\sqrt{2\pi\kappa_2}} \exp\left\{-\frac{D'^2}{2\kappa_2}\right\},$$

$$P_{D,D''}(D,D'') = \frac{1}{2\pi\sqrt{\kappa_0\kappa_4 - \kappa_2^2}} \exp\left\{-\frac{\left(D^2 \kappa_4 + D''^2 \kappa_0 + 2DD''\kappa_2\right)}{2\left(\kappa_0\kappa_4 - \kappa_2^2\right)}\right\}. \tag{A.11}$$

From this follows

$$P_Z(Z) = \int_{-\infty}^{\infty} dD \int_{-\infty}^{\infty} dD'' P_{D,D''}(D,D'') \delta(Z - DD'')$$

$$= \frac{1}{\pi\sqrt{\kappa_0\kappa_4 - \kappa_2^2}} \exp\left\{-\frac{Z\kappa_2}{(\kappa_0\kappa_4 - \kappa_2^2)}\right\} K_0\left(\frac{|Z|\sqrt{\kappa_0\kappa_4}}{\kappa_0\kappa_4 - \kappa_2^2}\right),$$
(A.12)

The statistics of $Q$ depend on a single parameter: the ratio

$$\alpha = \frac{\kappa_0\kappa_4}{\kappa_2^2},$$
(A.13)

which satisfies $1 \leq \alpha < \infty$. Defining the auxiliary quantities

$$a \equiv \frac{(\alpha+1)Q - 2}{(\alpha-1)Q}, \quad b \equiv \frac{|1-Q|\sqrt{\alpha}}{|Q|(\alpha-1)},$$
(A.14)

and using the $\delta$ function in (A.3) to eliminate $Z$, gives

$$P(Q) = \frac{2}{\pi^{3/2}\sqrt{2\kappa_2^{3/2}(\alpha-1)}Q^2} \int_0^{\infty} dD' D'^2 \exp\left\{-\frac{D'^2 a}{2\kappa_2}\right\} K_0\left\{\frac{D'^2 b}{\kappa_2}\right\}. \quad \text{(A.15)}$$

The integral can be evaluated in terms of elliptic integrals E and K, so, finally (and using the definitions in Mathematica$^{TM}$),

$$P(Q) = \frac{\left(-4b\mathrm{E}\left(\frac{1}{2} - \frac{a}{4b}\right) + (a+2b)\mathrm{K}\left(\frac{1}{2} - \frac{a}{4b}\right)\right)}{\pi\alpha^{1/4}\sqrt{|Q^3(1-Q)|(a^2 - 4b^2)}}.$$
(A.16)

Figure 6 shows the distributions for different values of the parameter $\alpha$. Large $\alpha$ corresponds to an ensemble of functions $D(x)$ whose Fourier components span a wide range of wavenumbers $\kappa$. Then there are many complex zeros and the distribution (e.g. figure 6a) has a large tail for negative $Q$ and so is very different from those we have studied in the main body of the paper. However, as $\alpha$ approaches 1 from above, the functions $D(x)$ become almost monochromatic, and in the limit we recover the simple example (1.9-1.11) (figure 6d).

**Figure captions**

**Figure 1**. (a) The function $Q(24,476,747+u)$ for the Riemann zeta function (1.3), computed using the Riemann-Siegel formula [14] with one correction term (after E O Tuck [1]). (b) $P(Q)$ for the Riemann zeta function, computed from 50000 samples over a stretch of the critical line containing 4216 zeros starting at $x$=251330, that is near the 383771st zero.

**Figure 2**. (a) Thick curve: $P(Q)$ for Poisson-distributed zeros, computed from 50000 samples of 20 randomly chosen zeros with the distribution (2.9); the largest value of $Q$ in these data was 10.61 – much smaller than the theoretical maximum $Q$=20. Dashed curve: large $N$ theoretical

distribution (5.6). Thin curve: asymptotic formula (5.13). (b) Thick curve: $P(Q)$ for GUE-distributed eigenvalues, computed from 50000 samples of 20x20 random Hermitian matrices, giving eigenvalues with the distribution (2.10); the largest value of $Q$ in these data was 2.66 – much smaller than the theoretical maximum $Q=20$. Dashed curve: $P(Q)$ for the Riemann zeros (figure 1b).

**Figure 3**. (a) Thick curve: $P_2(Q)$ for Poisson-distributed zeros, computed from 50000 samples of 2 randomly chosen zeros with the distribution (2.9). Dashed curve: theoretical distribution (3.14). (b) Thick curve: $P_2(Q)$ for GUE-distributed eigenvalues, computed from 50000 samples of 2x2 random Hermitian matrices giving eigenvalues with the distribution (2.10). Dashed curve: theoretical distribution (3.15).

**Figure 4**. (a) Thick curve: $P_3(Q)$ for Poisson-distributed zeros, computed from 50000 samples of 3 randomly chosen zeros with the distribution (2.9). Dashed curve: theoretical distribution (3.20). (b) Thick curve: $P_3(Q)$ for GUE-distributed eigenvalues, computed from 50000 samples of 3x3 random Hermitian matrices giving eigenvalues with the distribution (2.10). Dashed curve: theoretical distribution (3.22).

**Figure 5**. Phase contours of $\chi(\tau)$, defined by (5.3), intersecting at the zeros $\tau_j$ (poles of $\partial \log\chi(\tau)/\partial\tau$) given approximately by (5.9), and deformed integration contour for $Q>1$.

**Figure 6**. $P(Q)$ for gaussian random functions, for spectral parameters (a) $\alpha=20$, (b) $\alpha=5$, (c) $\alpha=2$, (d) $\alpha=1.001$.

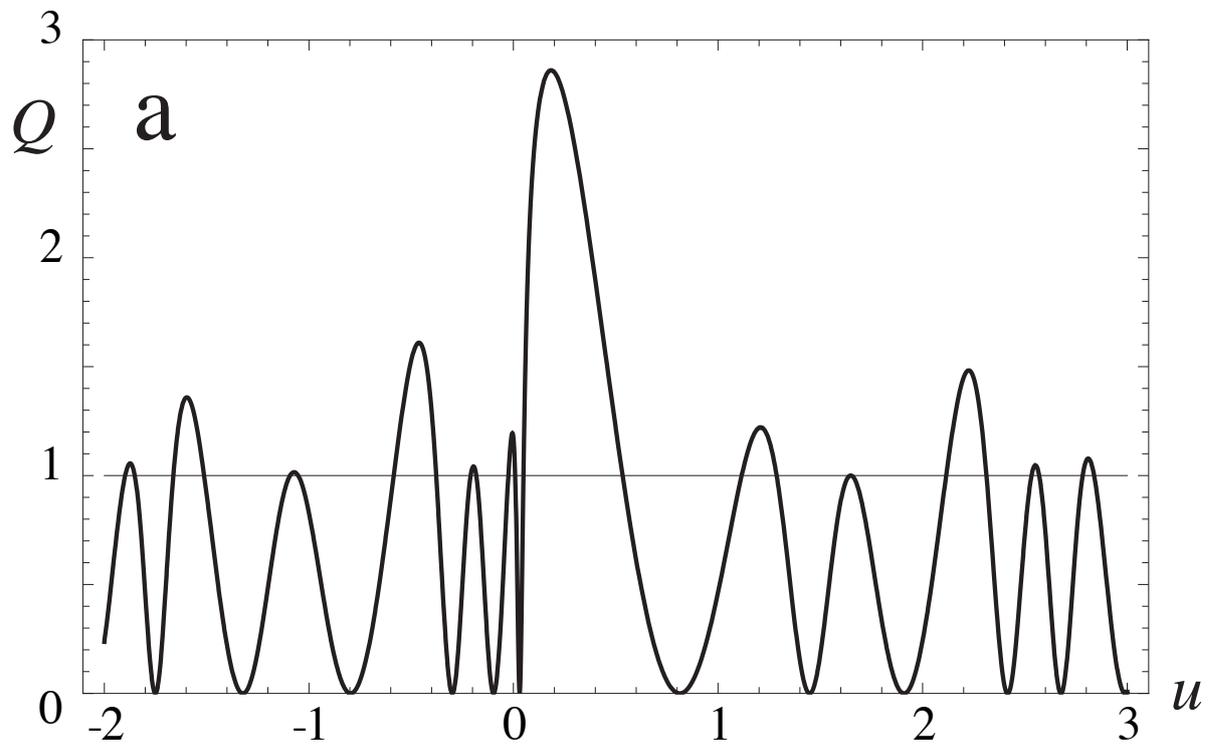

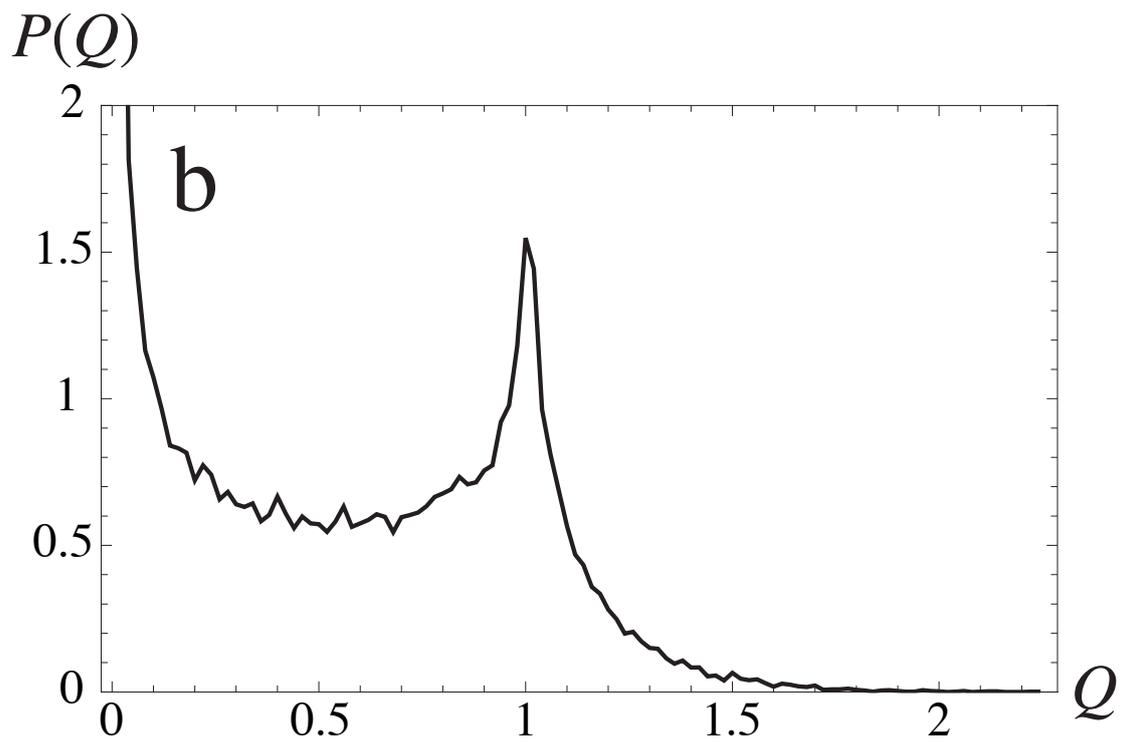

figure 1

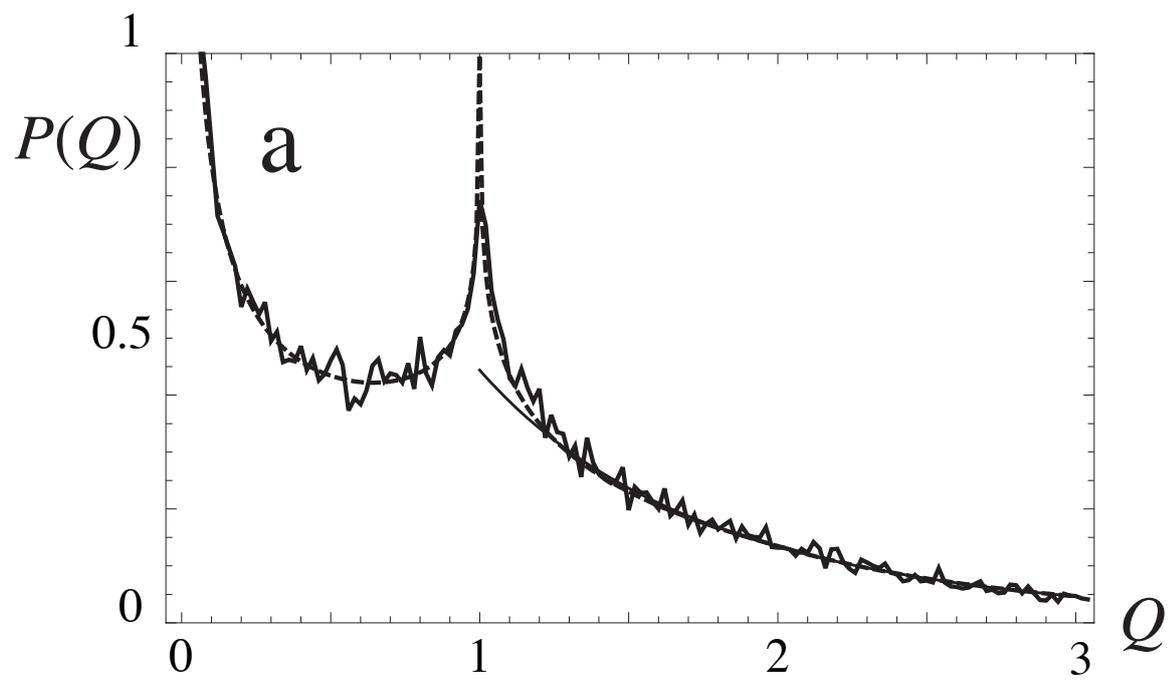

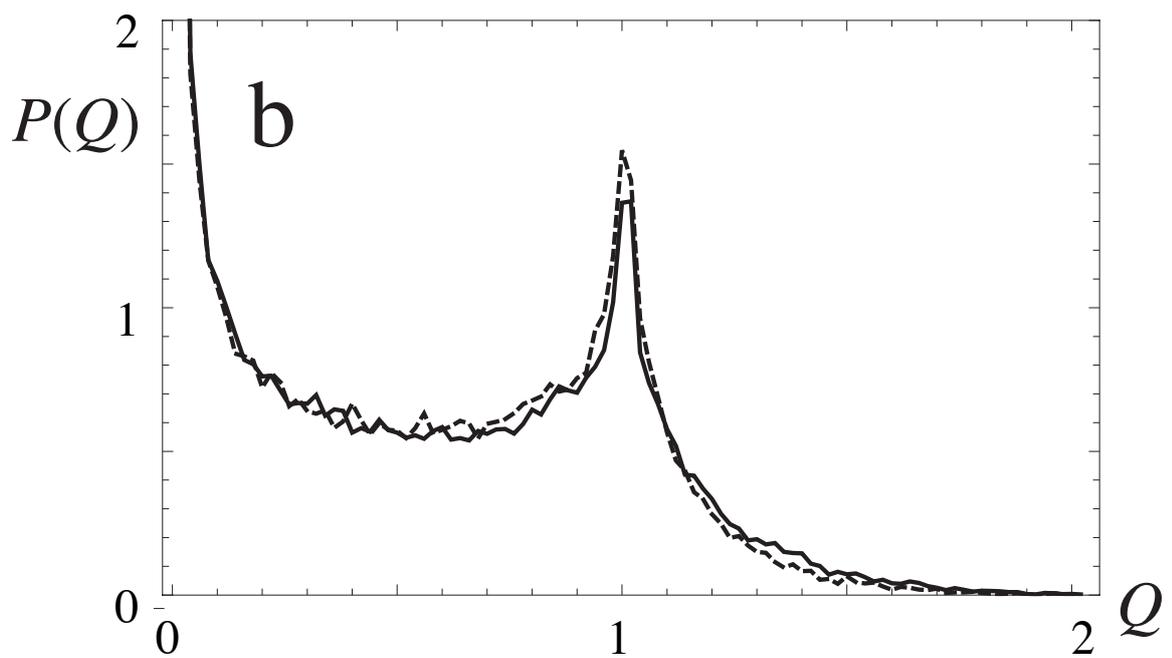

figure 2

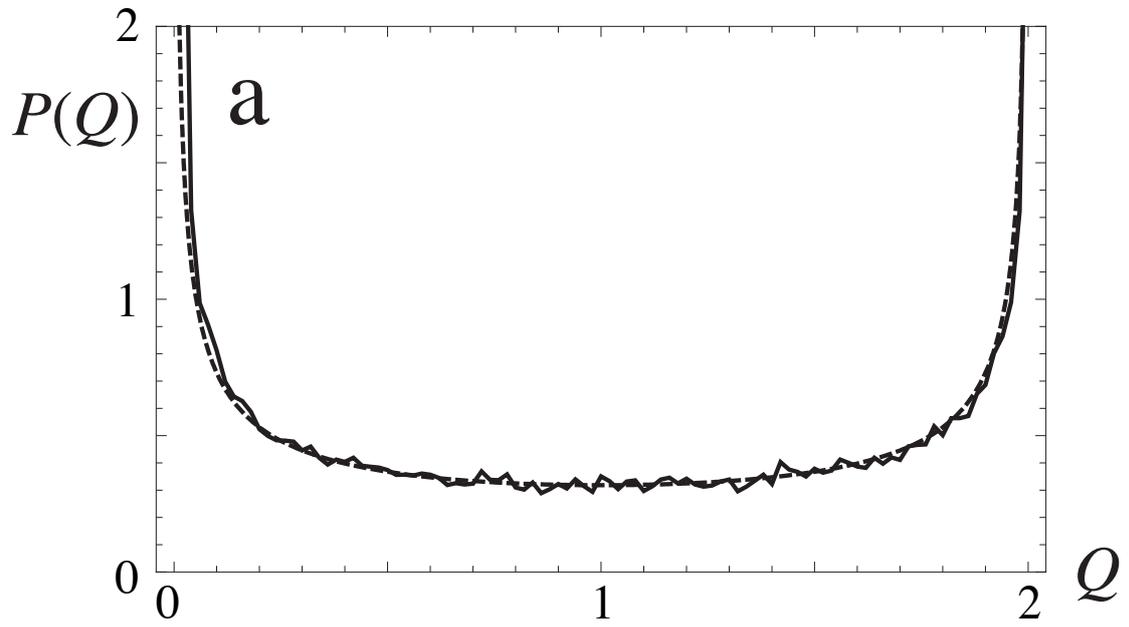

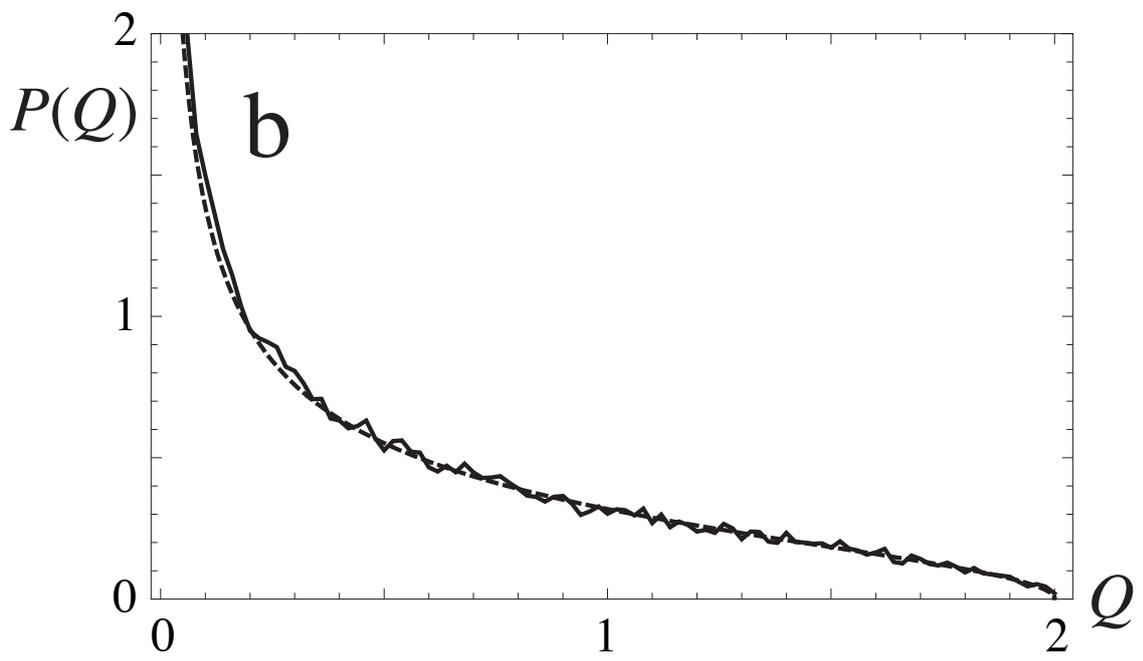

figure 3

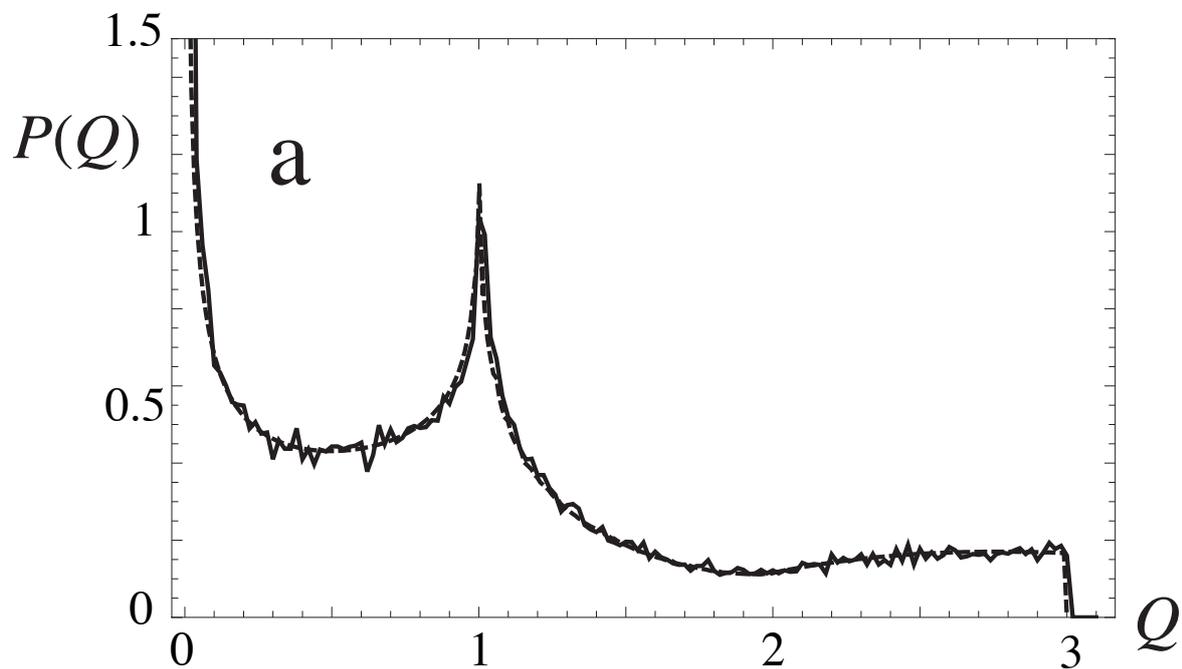

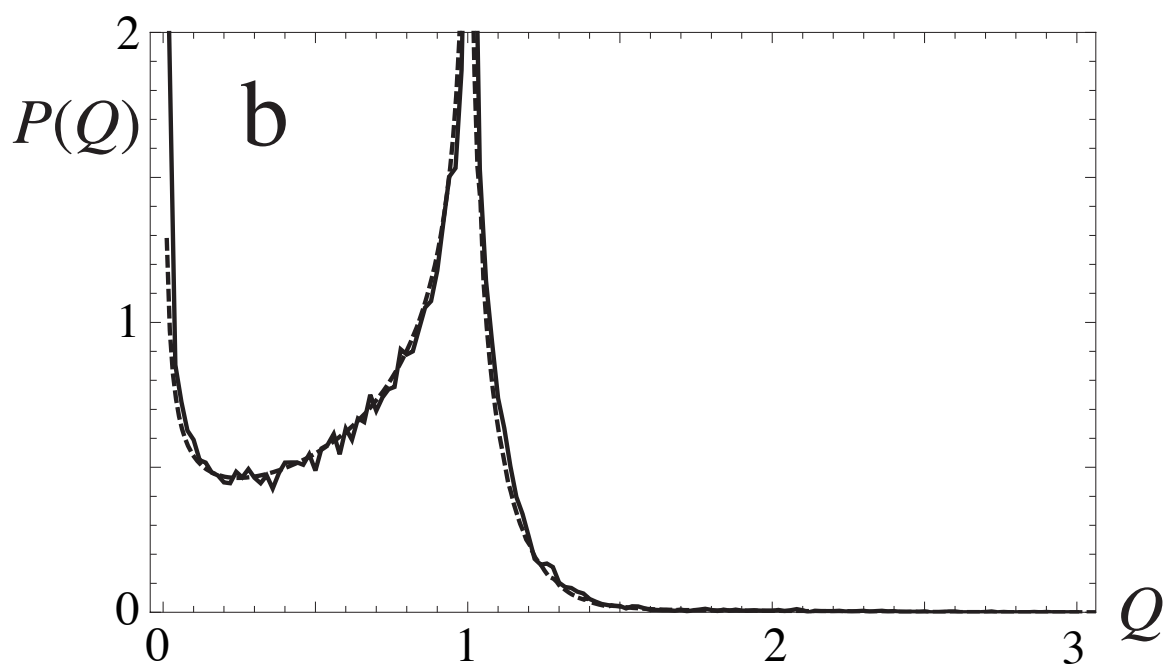

figure 4

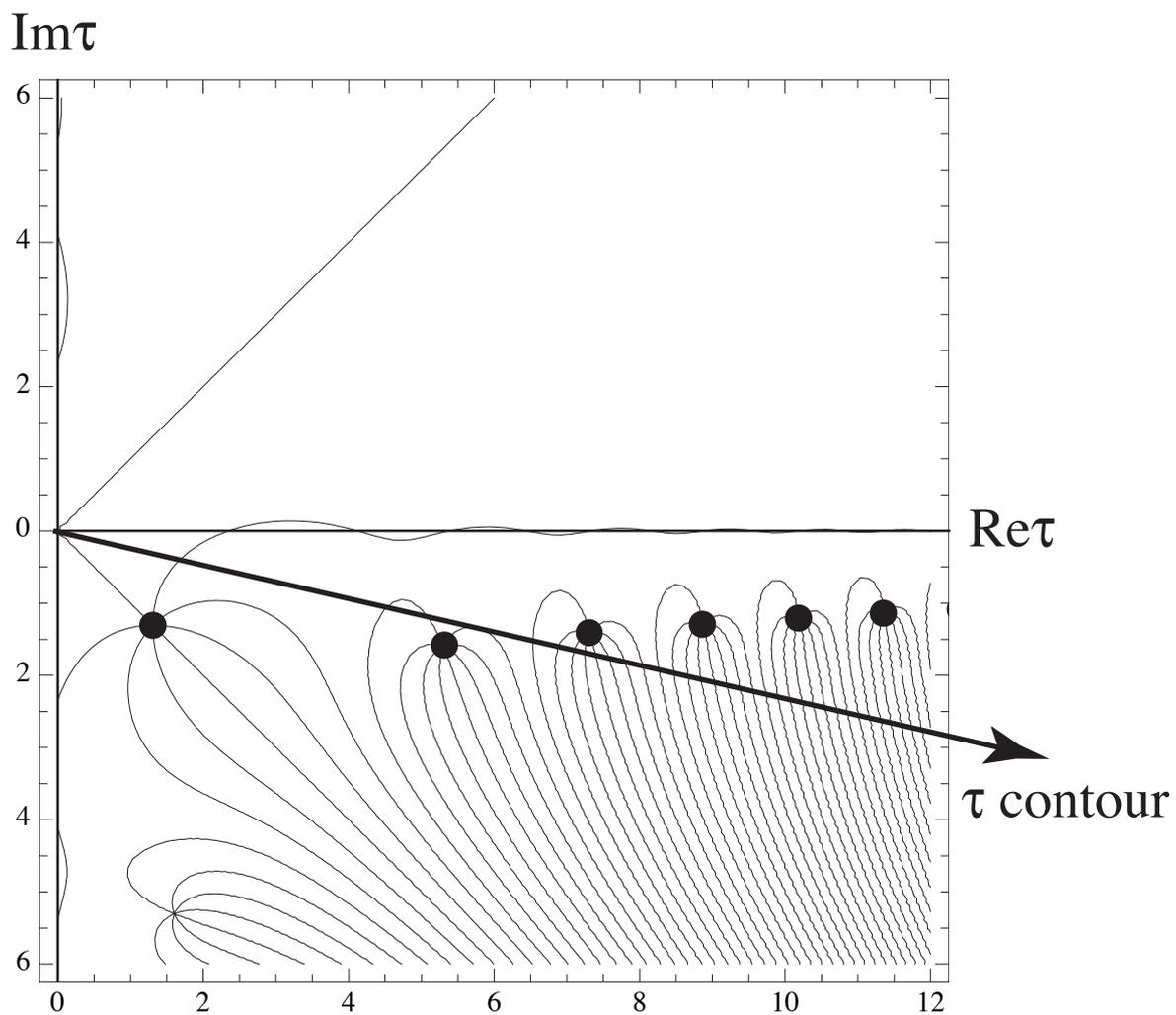

figure 5

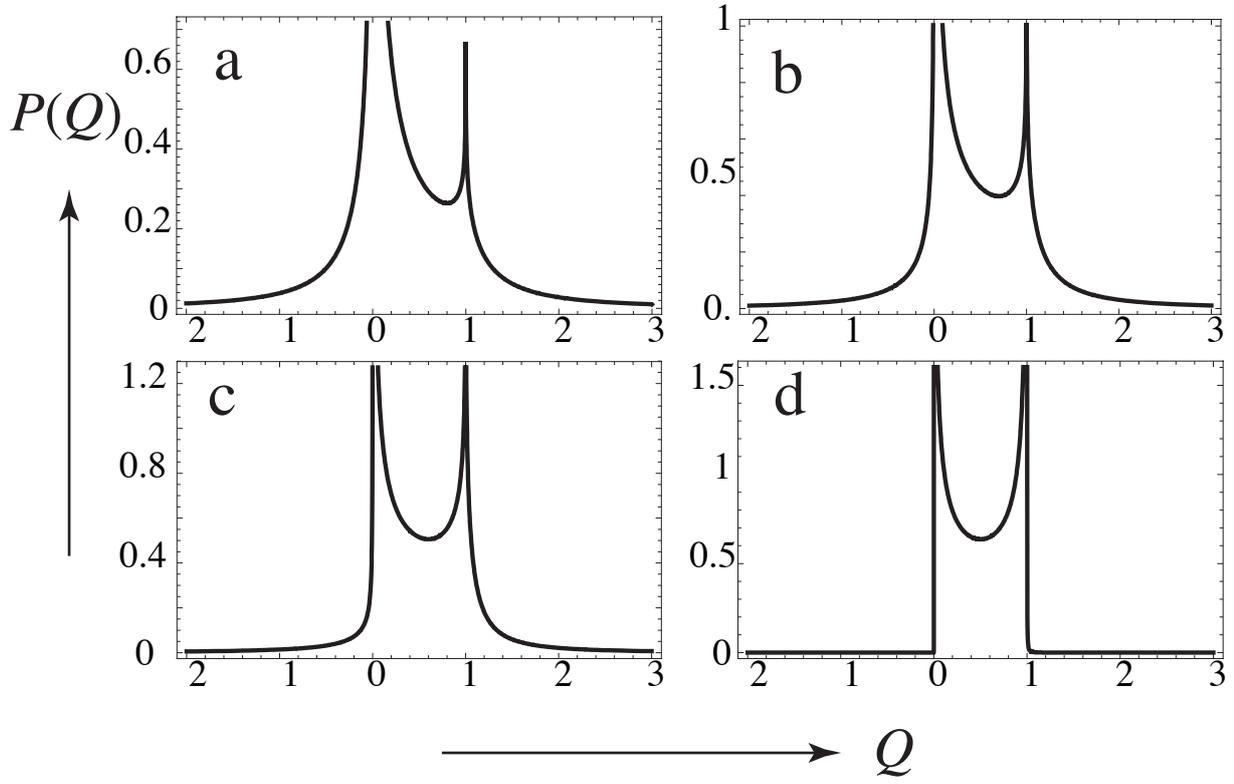

figure 6